\documentclass[12pt,a4paper]{article}
\usepackage[utf8]{inputenc}
\usepackage[francais]{babel}
\usepackage[T1]{fontenc}
\usepackage{lmodern}
\usepackage{amsmath}
\usepackage{amsfonts}
\usepackage{amssymb}

\usepackage{graphicx,}
\begin{document}

\title{Absorption spectrum of very low pressure atomic hydrogen.}

\author{Jacques Moret-Bailly\footnote{email: jmo@laposte.net}}

\maketitle

\begin{abstract}

Les spectres des quasars résultent essentiellement d'interactions de la lumière naturelle avec de l'hydrogène atomique. Une absorption visible d'une raie spectrale fine et saturée dans un gaz requiert une basse pression, donc un long parcours sans perturbation comme un rougissement cosmologique. Burbidge et Karlsson ont observé que les rougissements des spectres des quasars résultent de rougissements fondamentaux écrits 3K et 4K, qui amènent les raies d'absorption Lyman beta ou gamma de l'atome H sur la raie alpha.  Le spectre est donc rougi tant qu'une raie absorbée ne se superpose pas à la raie Lyman alpha du gaz: le rougissement se produit donc seulement si une absorption alpha pompe des atomes dans l'état 2P. L'espace est structuré en coquilles sphériques centrées sur le quasar contenant ou non des atomes excités. Dans les coquilles excitées la densité d'atomes 2P croît jusqu'à ce que l'amplification cohérente d'un mode ayant un long parcours dans la coquille, donc perpendiculaire au rayon observé, soit assez grande pour qu'un flash superradiant à la fréquence alpha éclate, absorbant l'énergie des atomes et du rayon observé, à la fréquence Lyman alpha locale. Ces oscillations de relaxation inscrivent de nombreuses raies d'absorption. Les rougissements des spectres des quasars sont dus à des interactions paramétriques composées d'effets Raman impulsionnels stimulés (ISRS): Des atomes d'hydrogène excités catalysent des échanges d'énergie entre un rayon observé et des rayons froids du fond thermique, en accord avec la thermodynamique.

\centerline
{\bf Abstract.}
Spectra of quasars result primarily from interactions of natural light with atomic hydrogen. A visible absorption of a sharp and saturated spectral line in a gas requires a low pressure, so a long path without blushing as a cosmological redshift. Burbidge and Karlsson observed that redshifts of quasars result from fundamental redshifts, written 3K and 4K, that cause a shift of absorbed beta and gamma lines of H to alpha gas line. Thus absorbed spectrum is shifted until an absorbed line overlaps with Lyman alpha line of gas: redshift only occurs if an alpha absorption pumps atoms to 2P state. Thus, space is divided into spherical shells centered on the quasar, containing or not 2P atoms. Neglecting  collisional de-excitations in absorbing shells, more and more atoms are excited until amplification of a beam having a long path in a shell, thus perpendicular to the observed ray, is large enough for a superradiant flash at alpha frequency. Energy is provided by atoms and observed ray, absorbing a line at local Lyman alpha frequency. Redshifts of quasars spectra are due to parametric interactions composed of Impulsive Stimulated Raman Scatterings (ISRS) : Excited hydrogen atoms catalyze energy exchanges between observed light rays and background cold rays, in agreement with thermodynamics.

\end{abstract}
Radiative transfer.
Scattering.

98.62.Ra Intergalactic matter; quasar absorption and emission-line systems; Lyman forest

290.5910 Scatte
ring, stimulated Raman 

190.2640 Nonlinear optics : Stimulated scattering, modulation, etc.

\section{Introduction.}
\label{Intro}
Study of spectra of quasars is difficult because usual spectroscopy which uses transitions between energy levels does not work.

\medskip
Burbidge \cite{Burbidge}, then Karlsson \cite{Karlsson} found an empirical formula on redshifts of selected quasars, formula which is transformed in section \ref{disc}, so that its empirical parameter (Karlsson's constant) appears connected to Lyman spectrum of H atom.

Subsection \ref{mod1} applies transformed formula to spectral frequencies, correlating them. It appears that light is redshifted when Lyman $\alpha$ line is absorbed, while visible absorption of gas lines requires a stop of redshift which occurs if there remains no energy at Lyman $\alpha$ frequency.

Remarkable redshifts 3K or 4K (where K is Karlsson's constant 0.061) show that redshifts stop if absorbed lines Ly$_\beta$ or Ly$_\gamma$  reach $\nu_\alpha$ frequency. Thus we suppose that atoms in state 2P produce a redshift, so that  a lack of these atoms allows an absorption of intense, sharp lines.

In \ref{mod1}, these remarks are used to build a set of spectral lines.

In \ref{struc}, space is divided into spherical shells around star, absorbing (thus redshifting)  or not light at local Lyman $\alpha$ frequency.

 In section \ref{flash}, gas in a redshifting shell may lose, during a short time, its excitation by emission of a superradiant beam, so that a sharp absorbed line appears.

Section \ref{isrs}, shows how impulsive stimulated Raman scatterings (ISRS) redshifts light.

\section{Absorption of a gas spectrum by lack of energy at Lyman $\nu_\alpha$ frequency of H atom. }

\subsection{Karlsson's and Burbidge's periodicities.}\label{peri}
Burbidge \cite{Burbidge}, then Karlsson \cite{Karlsson} observed that relative frequency shifts of spectral lines of quasars z = [$\nu_{emit}-\nu_{obsv}] / \nu_{obsv}$ are often close to remarkable values $z = 0.061$ and $z=1.95$. This result was generalized by Karlsson's formula:

 $z(n)= nK$, (1)
 
 where n is integer 3, 4, 6,... and K=0.061. 

Many sharp, saturated lines  making "Lyman forests" of quasar obey these rules; many authors tried to apply similar formula to galaxies, but it failed. 
  
As his formula does not work well for large $n$, Karlsson proposed a periodicity of ln(1+z), additions of which multiplies sets of frequencies by parameters, as Doppler redshifts do, without very good results.

An aim of this paper is extension of Karlsson's work to understand formation of \textquotedblleft Lyman forests" of quasars. We use mainly conventions and results found in P. Petitjean's paper \cite{Petitjean}, also Rauch's ideas \cite{Rauch}.

\subsection{Discussion of Karlsson's formula.}
\label{disc}

In Doppler frequency shifts, emitted and received frequencies verify equation which involves, as parameters, emission and reception speeds $V_{emission}$ and $V_{reception}$ :

 $\nu_{reception}=\frac{c-V_{reception}}{c-V_{emission}}*\nu_{emission}$	(2)
 
 We assume that frequency shifts at various light frequencies obey an equation similar to equation (2), so that a single parameter is needed to compute frequency shifts at all frequencies, for instance the ratio $f_1/f_0$ of a particular shifted frequency over initial frequency :

 $\nu_{shifted}=\frac{f_1}{f_0}\nu_{initial}$ 	(3)

The terms of strange serie 3, 4, 6, ... of Karlsson's formula may be written $3p+4q$, where $p$ and $q$ are any non-negative integers. Thus redshifts $3K$ and $4K$ are remarkable, able to generate simply all Karlsson's redshifts.

Compute redshifts which put Lyman $\nu_\beta$ and $\nu_\gamma$ frequencies of H atom to  $\nu_\alpha$ frequency:

$Z_{(\beta,\alpha)} = (\nu_\beta-\nu_\alpha)/\nu_\alpha = [(1-1/32 -(1-1/22)]/(1-1/22) ] \approx 5/27 \approx 0.1852 \approx 3*0.0617 \approx 3*K; $ (4)
 	
$Z_{(\gamma,\alpha)} = (\nu_\gamma-\nu_\alpha)/\nu_\alpha = $[(1-1/42 -(1-1/22)]/(1-1/22) ] = 1/4 = 0,25 = 4*0.0625 $\approx 4K$. (5)

Assuming that use of  3 lowest Lyman frequencies of H atom is better than use of K, and that variation of shifts at various frequencies is obtained from a particular shift as in a Doppler redshift,  Karlsson's formula becomes simpler: shifted frequencies $\nu(p,q)$ depend only on absolute frequency $\nu_0$ , three well known frequencies and two any non-negative integers $p$ and $q$:

$\nu(p,q) = (\nu_\alpha/\nu_\beta)^p * (\nu_\alpha/\nu_\gamma)^q * \nu_0$.	(6)
 
Karlsson's formula (1), applied, for instance, to Lyman beta line with n=3 does not generate exactly Lyman alpha frequency, while formula (6) does. Though resolution of spectra does not allow to distinguish a doublet from a superposition of lines, we will choose formula (6) because choice of possible values of $p$ and $q$ seems more natural than extraction of $n$ from a less natural serie. Is it only aesthetics ?

\subsection{Model of a quasar surrounded exclusively by hydrogen.}
\label{mod1}
\subsubsection{Possible hypothesis.}

We assume that ``Lyman forest'' is built by absorption of a thermal emission of an extremely hot star by {\it pure}, relatively cold (2 000-50 000 K), low pressure, non-excited, atomic hydrogen.

Following properties of many lines of ``Lyman forest''  are directly deduced from Petitjean's paper \cite{Petitjean}:

-A- As lines are sharp, widening of lines by collisions must be negligible, pressure of gas must be very low.

-B- To obtain absorption of Lyman lines at several frequencies, a redshift process of electromagnetic waves is necessary. Here, we make only hypothesis of formula (3) about this redshift.

-C- A single, unshifted Ly$_\beta$ is observed; no unshifted Ly$_\gamma$ appears because it does not remain absorbable energy at high frequencies after redshifts of thermal emission profile. 

\medskip
Absence of an absorption line may result from :
  
-a- Absence of emitted energy around frequency of line.

-b- A permanent shift of light frequencies dilutes absorption, so that all absorbed (or emitted) lines have width of shift and lines are weak, not observable. Accordingly, absorption (or emission) of sharp lines requires a stop of frequency shifts.

-c- Accurate superposition of observed lines results from the choice of redshift equations:
Absence of shifted $\beta$ and $\gamma$ lines while sharp $\alpha$ lines are observed, is due to superposition of these lines with $\alpha$ lines.

\medskip
In standard theory, sharpness of saturated absorbed lines requires contradictory conditions: 

- Column density of gas must be large for saturation;

- Absorbing gas must be thin to avoid a broadening of lines by frequency shift during absorption.

- Pressure of gas must be low to avoid collisional broadening of line.

Thus gas must be in filaments which are only detected on paths from quasars.

\medskip
Supposing that redshift is related to a physical property of gas, condition is:

- Light is redshifted except if an absorbed line is at Lyman $\alpha$ frequency. Thus redshift requires a lyman  $\alpha$ absorption, that is generation of 2P atomic hydrogen.

\subsubsection{Building of spectrum.}

Figure 1 represents a canvas of atomic hydrogen spectrum for building any absorption spectrum by addition of lines, in particular able to play the role of $Ly_\beta$ or $Ly_\gamma$ lines.
\begin {figure*}

\label{spectre}
\centering
\includegraphics[width=16cm]{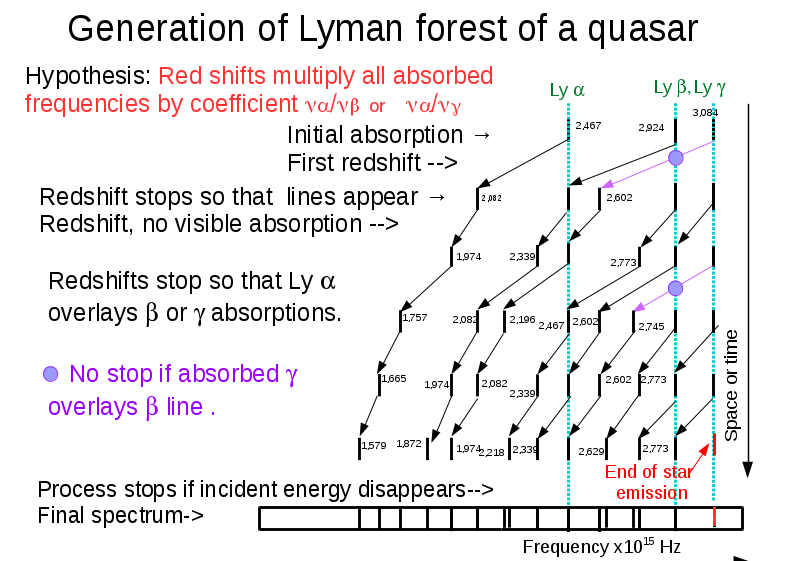}

\caption{Generation of lines of Lyman forest of a quasar by coincidence of lines $\alpha$ with more shifted $\beta$ or $\gamma$. During stop of redshift Ly$_\beta$ and Ly$_\gamma$ lines are absorbed. Lines of other local gas may be also absorbed and may later play the role of Ly$_\beta$ and Ly$_\gamma$ lines if their frequencies are larger than $\nu_\alpha$. Written frequencies do not take into account dispersion of hyperfine polarizability of H atom.}
\end{figure*}

Rules used to build spectrum taking into account only hydrogen atoms are simple :

  - i - At start, close to star, we suppose that $\alpha, \beta$ and $\gamma$ lines have been absorbed.
  
  - ii - It appears a frequency shift until an absorbed, shifted line of initial frequency $\nu$ reaches $\nu_\alpha$ frequency. Thus all absorbed frequencies have been multiplied by $\nu_\alpha/\nu$, coefficent lower than 1. In pure atomic hydrogen $\nu$ may be $\nu_\beta$ or $\nu_\gamma$, assuming that higher frequency lines are too weak.
  
  - iii - During stop of frequency shift, the three main lines of H could be absorbed, but there remain no energy at Ly$_\alpha$ frequency. 
  
   - iv - Assume that Ly$_\beta$ absorption produces a very weak redshift. During this negligible redshift, gas lines are visibly absorbed. If weak redshift is able to shift absorbed frequencies off Ly$_\alpha$ absorption line before full absorption of light at Ly$_\beta$ frequency, fast redshift restarts, go to - ii -. Else there is no more redshift, absorbed Ly$_\beta$ line is visible, but Ly$_\gamma$ line is probably not because its frequency is probably larger than the shifted high frequency limit  of emission of star. 
   
   In pure H, total redshift results from several relative shifts Z$_{\beta, \alpha}$ and  Z$_{\gamma,\alpha}$ which correspond to multiplication of light frequencies by $\nu_\alpha/\nu_\beta$ or $\nu_\alpha/\nu_\gamma$. 
   
Computed lower frequencies written on figure are not very good. They can be corrected by multiplication by a unique dispersion function $F(\nu)$ equal to 1 at $\nu_\alpha$ frequency.

\subsubsection{Introduction of other gas in spectrum building.}
Lines found in figure 1 appear in real spectra, after a correction of their frequencies by the factor  $F(\nu)$. But many lines have an other origin: various lines absorbed at frequencies larger than $\nu_\alpha$ may play the role of Ly$_\beta$ and Ly$_\gamma$, multiplying density of lines.

Figures much more complex than fig. 1 can be drawn.

\subsubsection{Structuring space.}
\label{struc}
Redshifts stop if frequency of an absorbed line becomes $\nu_\alpha$ and not if absorbed $\nu_\gamma$ frequency reaches $\nu_\beta$ frequency. This shows that redshift results on generation of 2P hydrogen atoms.

Up to now, we have only supposed that it exists somewhere (sometime for light) a redshift. Thus, assuming that quasar is far from other stars, space is divided into spherical shells where generated 2P atoms shift light frequencies, and shells in which there is no 2P atoms and no frequency shifts.

As quasars are not alone in space, perturbation of generation of these regions may result from pumpings of atoms by light of other stars. Thus region in which quantized redshifts appear must be small enough to avoid an important lighting by other stars: This explains Burbidge's selection of quasars.

To avoid different pumpings by light coming from different regions of surface of the star, redshift must appear only at a distance much larger than size of the star, that is in a region where pressure of gas is low.

These conditions are evidently not verified for a set of relatively close stars as a galaxy. As the star must produce very high light frequencies, if it is small it must be extremely hot. An hypothesis which seems valuable is an accreting neutron star, a type of stars which were never found in nebulae while they should be seen.

\section{Absorption of a gas spectrum by lack of energy at Lyman $\nu_\alpha$ frequency of H atom: Dynamics of space structure.}
\label{flash}
In model described in \ref{mod1}, we obtained spherical shells of hydrogen atoms either excited by absorption of Lyman $\nu_\alpha$ frequency, or in ground state. But it is difficult to obtain the large number of lines observed for instance between Ly$_\alpha$ and  Ly$_\beta$ lines ( Ly$_\alpha$ forest ).

In \ref{struc}, we took only ito account the lack of absorption at $\nu_\alpha$ frequency to explain stops of redshifts. But, in spherical shells in which $\nu_\alpha$ frequency is absorbed, Einstein's B coefficient of amplification at $\nu_\alpha$ may become large because temperature of exciting light is large and there are almost no collisions to de-excite atoms. Thus, Lyman alpha coherent, superradiant flash emissions may burst, into directions almost orthogonal to exciting rays, for which path in a spherical shell is maximum. 
The flash pumps energy from atoms and light which excites them, so that an absorption line appears in spectrum. These transfers of energy are in agreement with thermodynamics because energy flows from hot beams comming from star to resulting beams stemming from dark infinity through flash, radiance of which is average. The result is a sharp absorption at current frequency .

It is difficult to discuss on stability of relaxation oscillations which generate these absorptions, thus on stability of spectra. However, the medium is probably much more homogeneous and stable than the medium which provides aurora borealis, similar to permanently superradiant media which provide punctuated or not circles around Strömgren spheres, possibly transformed by planets into hourglasses as in SNR1987A. A computation of this spectrum seems very difficult, requiring a lot of hypothesis.

\section{Physical interpretation.}
\label{isrs}
\subsection{Comparison of optical properties of 1S and 2P hydrogen.}
In 1S state, hydrogen has 1420 MHZ hyperfine frequency, (period T=0.7ns;  wavelength $\lambda$ =  21 cm),  quadrupolar resonance frequency.

In its first excited state, hydrogen atom has quadrupolar resonance frequencies: 178 MHz (T=5.6 ns, $\lambda$ =1,7 m) in state 2S$_{1/2}$, 59 Mhz (T=17 ns,  $\lambda$ = 5 m) in  2P$_{1/2}$ and 24 MHz (T=42 ns, $\lambda$ =  12 m) in 2P$_{3/2}$.

A qualitative difference between laboratory and space ISRS is that ordinary incoherent light is made of longer, around 1 nanosecond  pulses: these pulses must be shorter than hyperfine periods in excited states, as required by conditions of space coherence of "Impulsive Stimulated Raman Scattering" (ISRS) : {\it Length of pulses must be shorter than all involved time constants} \cite{GLamb}.

Space coherence of incident and scattered light in ISRS allows an interference of exciting and scattered light, which shifts frequency of incident light and preserves the geometry of light beams.
 
The other condition for ISRS is: Collisional time must be longer than 1ns, that is pressure must be very low, so that pressure broadening of lines is low.

These conditions make ISRS in space very weak compared with ISRS in labs of chemistry, which uses around  10 nanosecond laser pulses, around $k=10^5$ times shorter than light pulses of temporally incoherent light:

- Division of pressure by k divides the shift by k.

- Division of quadrupolar frequency by k divides the shift by k twice:

  - - Once mixing incident and Raman frequency shifted of hyperfine frequency;
 
  - - Once by division of difference of populations of hyperfine levels, assuming thermal equilibrium.
  
  Thus order of magnitude of path needed to observe ISRS is $\approx 10^{15}$ times longer than in laboratory experiments: astronomical paths are needed for observation.
  
\subsection{De-excitation of hyperfine levels.}
As pressure is very low, collisions between atoms are negligible. A radiative process is necessary to de-excite hyperfine levels, thus obtain Raman frequency shift.

Happily, there are cold background electromagnetic waves. Thus, the real process is not a single ISRS, but a set of ISRS such that hyperfine levels of 2P atoms are not excited. Excited atoms H catalyze an exchange of energy between light beams, it is a {\it parametric interaction}. Variations of energy produce frequency shifts of light beams, so that their entropy is increased. This parametric interaction is usually named "Coherent Raman Effects between Incoherent Light beams" (CREIL).

\section{Conclusion.}
Study of Lyman forests of quasars is difficult because they mix spectra origins of which are different, although all visibly absorbed lines are sharp and saturated :

- Karlsson's relative frequency shifts 3K or 4K transform, shift, with a good approximation, absorbed Lyman frequencies $\nu_\beta$ or $\nu_\gamma$ of H atom into $\nu_\alpha$ frequency. During corresponding frequency shifts, H atoms are pumped to 2P level by permanent, diluted, invisible $Ly_\alpha$ absorption.

- Fast, superradiant de-excitation of 2P atoms by emission of a superradiant flash absorbs light at $\nu_\alpha$ frequency, during a relaxation process. Burbidge and Karlsson selected quasars in which few lines have this origin to obtain their results.

Coherently scattered Raman light interferes with incident light, shifting its frequency (As Rayleigh scattering shifts the phase, producing refraction).

Stop of redshift which allows saturated, visible absorption of lines is due to lack of 2P atoms.
This lack may result from flash, coherent de-excitation of 2P levels, or to presence of an absorbed line at  $Ly_\alpha$ frequency. In this last case, redshift may remain very weak by $Ly_\beta$ absorption, until absorbed line is shifted out of $Ly_\alpha$ frequency, so that fast redshift restarts...

Additional absorptions by various atoms may also play the role of Lyman $\beta$ and $\gamma$ absorptions, complicating  the spectrum. Comparison of theoretical and observed spectra shows a small shift increasing at lowest frequencies, because, as refraction, ISRS has a dispersion which was not taken into account. 

Hubble's law does not apply to quasars which may be accreting neutron stars lying in close galaxies.

\end{document}